\documentclass[twocolumn,preprintnumbers,amsmath,amssymb,aps]{revtex4}

\usepackage{graphicx}
\usepackage{amssymb}
\usepackage{epstopdf}
\usepackage{bm}

\begin{document}
    
\title{Many-Worlds Quantum Mechanics is Neither Mathematically Nor Experimentally Equivalent to Standard Quantum Mechanics}
    
\author{Frank  J. Tipler}
\affiliation{Department of Mathematics, Tulane University, New Orleans, LA 70118}

\date{\today}

\begin{abstract}
Many-Worlds quantum mechanics differs from standard quantum mechanics in that in Many-Worlds, the wave function is a relative density of universes in the multiverse amplitude rather than a probability amplitude.  This means that in Many-Worlds, the Born frequencies are approached rather than given {\it a priori}.  Thus in Many-Worlds the rate of approach to the final frequencies can be calculated and compared with observation.  I use Many-Worlds to derive the rate of approach in the double slit experiment, and show that it agrees with observation.  Standard quantum theory has never been used to derive an approach formula because it cannot be so used, as has been tacitly acknowledged for 70 years.
\end{abstract}

\maketitle

Standard quantum mechanics and Many-Worlds quantum mechanics are neither mathematically nor experimentally equivalent because they differ on the meaning of the wave function $\psi$. Let me show this by deriving the Schr\"odinger equation from classical mechanics.  I shall show that the Schr\"odoger equation is nothing but the mathematically consistent form of the general Hamilton-Jacobi equation, which is:

\begin{equation}
\frac{\partial S}{\partial t} + \sum_{i=1}^N\frac{1}{2m_i}(\vec\nabla S)^2 + V(\vec x, t) = 0
\label{eq:HJequation}
\end{equation}

\noindent
where $S$ is the action. The vector $\vec x$ refers all the coordinates of all the particles, there being $n_i$ particles for each mass $m_i$.   A particle with mass $m_i$ has a configuration space trajectory tangent to $\vec\nabla_i S/m_i$.  There are an uncountable infinity of such trajectories.

In standard H-J theory, only one such trajectory is held to exist, but the equation asserts the existence of all the trajectories, as emphasized in 1891 by the great mathematician Felix Klein \cite{KleinFelix1891}.  Taking the equation seriously means that a relative number density $R^2$ of trajectories can be defined, squared to ensure number density is non-negative, and ``relative'' because uncountability means if we label a countable number, we can change the number labeled by any factor $a$ without changing the physics.   Disappearance of trajectories would mean determinism breakdown at the disappearance point.  Thus, the density of trajectories must be conserved:

\begin{equation}
\frac{\partial{\rm R}^2}{\partial t} +  \sum_{i=1}^N\vec\nabla_i\cdot
\left({\rm R}^2\frac{{\vec\nabla_i}  S}{m_i}\right)=0
\label{eq:continuityequation}
\end{equation}

\noindent
Note that (\ref{eq:continuityequation}) is invariant under $R^2 \rightarrow aR^2$.

If the potential is attractive, the trajectories will bend toward one another on each side of the potential center, causing a breakdown in both (\ref	{eq:HJequation}) and (\ref{eq:continuityequation}).  Maxwell solved a similar equation breakdown by adding a term; we should do the same.  The derivative in (\ref{eq:HJequation}) causes the singularity, so try subtracting $\nabla_i^2R$ (linear in $R$, otherwise we get another nonlinearlty; subtract to cancel).  We must however enforce density relativity by dividing by $R$:  $-\nabla_i^2R/R$.  To make sure this cancels the problem in the derivative, generalize to $ -\sum_{i=1}^N(1/2m_i)\nabla_i^2R/R$.  Finally, get the units consistent by multiplying by a constant, call it $\hbar^2$, with units of action squared.  The proposed term to add to the potential in (\ref{eq:HJequation}) is thus $- \sum_{i=1}^N(\hbar^2/2m_i)\nabla_i^2R/R$ 

Defining $\psi \equiv Re^{iS/\hbar}$, the two equations (\ref{eq:HJequation}) with the new term and (\ref{eq:continuityequation}) can be written as a single equation:

\begin{equation}
i\hbar\frac{\partial\psi}{ \partial t} + \sum_{i=1}^N\frac{\hbar^2}{
2m_i}{\nabla^2_i\psi} - V(\vec x, t)\psi = 0
 \label{eq:QM}
\end{equation}

\noindent
which, because it is linear, has no singularities.  Equation (\ref{eq:QM}) thus is the mathematically consistent form of the Hamilton-Jacobi equation.  It is of course the Schr\"odinger equation.  Notice however that the above derivation leaves us no choice in choosing the meaning of $\psi$: it is necessarily a relative density of trajectories (worlds or universes) amplitude, not a probability amplitude.  Similarly, the derivation tells us what 
$\hbar$ is: the strength of interaction between the worlds.  It also tells us what a ``classical'' world is:  a trajectory along which the other worlds can be ignored.  This can occur only if $\nabla^2R(t) = 0$, which has $R(t) = {\rm constant}$ (independent of space) as the only bounded regular solution. (The so-called ``classical limit'' defined by $\hbar \rightarrow 0$ cannot define a classical system, because after May 20, 2019, mass is defined by $\hbar$, and thus $\hbar \rightarrow 0$ will imply $m\rightarrow0$.)  If $V(\vec x, t) =0$, the Many-Worlds classical trajectories are the plane waves, not minimal uncertainty wave packets.  Many-Worlds quantum mechanics is thus more general that standard quantum mechanics, because the wave function need not be an element in a Hilbert space.  I have elsewhere {\cite{Tipler2005} shown why the universe contains such classical states.  Finally, the derivation tells us {\it why} the equations of physics are quantized: to prevent singularities from arising in the laboratory, and this in turn tells us that linearity is essential, which means that Feynman sum-over-histories is the appropriate method for quantizing relativistic fields.  

The Many-Worlds meaning of $\psi$ entails that the Born frequencies are approached in the limit of an infinite number of experiments. All proponents agree that Born frequencies are the limit (e.g., \cite{Everett1957}, \cite{DeWittGraham1973}, \cite{Weinberg2015}), but they disagree on the physics underlying the approach.  The point which all previous analysts have missed is the the fact that a measurement is fundamentally an interaction of a {\it quantum system}, a system for which $\hbar$ cannot be ignored, with a {\it classical system}, a system for which $R(t) = {\rm constant}$.  It is the classical system, not the quantum system, that is the ultimate source of quantum probability, which is probability in the Laplace sense.  To illustrate Laplacean probability, consider a standard six-sided die.  To determine the probability that a given side will turn up, Laplace said , because the six sides are indistinguishable,  each side must have the same probability.  Since the probabilities must sum to 1, the probability that a given side will come up is therefore 1/6.  These probabilities can ``induce'' probabilities on interacting subsystems.  For example, they imply that the probability of a composite number appearing on a die is 1/3.  In the measurement process, it is the classical ``observers,'' --- any device for which $\hbar$ can be ignored --- that are indistinguishable (in the {\it quantum} sense of ``indistinguishable'') before the measurement.  Since for the classical observers, $R(t)={\rm constant}$, each possible observation is equally probable, in the Laplace sense.  Thus, at each observation, like a composite number on a die, the number of classical observers who measure a possible value will equal the relative density of the corresponding value in the multiverse of the quantum system.  So, if the classical observers are measuring the up or down spin of an electron, and the electron is the state $|\psi> = \sqrt{\frac{2}{3}}|\uparrow> + \sqrt{\frac{1}{3}}|\downarrow>$, then two-thirds of the classical observers will measure the spin to be up and one third will measure the spin to the down.  In Laplace probability theory, indistinguishability requires that we assign probability 2/3 to measuring spin up, and probability 1/3 to measuring spin down.  Relative probabilities can be used to calculate  \cite{Jaynes03} most probable expected frequencies, and the most probable frequencies of the spins equal the probabilities 2/3 and 1/3.  I have shown \cite{Tipler2014} that measured frequencies approach the Born frequencies with probability 1 in the limit of an infinity of measurements.  In the interaction of the $R(t)={\rm constant}$ observers with a quantum system, seeing a datum and not seeing a datum constitutes a Bernoulli process \cite{Tipler2014}.  This new type of probability arising from the interaction of a classical system with a quantum system, defines a new quantum statistics, call it Gibbs-Tipler statistics, the latter name suggested by ``drunkard's walk.''  But this new statistics shows quantum mechanics is only apparently indeterministic: equation (\ref{eq:QM}) is {\it more} deterministic than (\ref{eq:HJequation}).  

Einstein was correct: God does not play dice with the universe.

The historian Paul Forman was also correct that physicists in the 1920's imposed indeterminism on physics for philosophical reasons \cite{Forman1971}.  Thinking of the wave function as a probability amplitude was always incoherent, since it confuses probabilities with frequencies (\cite{Jaynes03}, especially Chapter 9).  The Born frequencies are indeed the limiting measured frequencies, as I showed in \cite{Tipler2014}, and restricting attention to the limiting case was sufficient as long as one restricted attention to huge numbers of atoms or electrons.  It is not sufficient when one measures individual electrons, as is done in the double slit experiment.

The fact that the frequencies are approached rather than assumed as a postulate is yet another way in which Many-Worlds quantum mechanics is more general than standard quantum mechanics.  

The rate of convergence to the final Born frequency pattern in any process, such as a Bernoulli process, which has a finite absolute third moment is given by the Berry-Esseen Theorem (\cite{Feller1972}, \cite{KorolevShevtsova2010}).  In the quantum case, the Berry-Esseen Theorem becomes in the case of one dimension:

\begin{eqnarray}
\left|\left[\frac{\sum_{i=1}^{j(x)} N_i}{N}\right] - \left[\frac{\int_a^x |\psi(t)|^2\,dt}{\int_a^b |\psi(t)|^2\,dt}\right] \right| \, \leq \,  \left(\frac{3+\sqrt{10}}{6\sqrt{2\pi}}\right)\times  \nonumber\\ 
\times\left[\frac{\left(\int_c^d|t|^3|\psi(t)|^2\,dt\right)\left(\int_c^d |\psi(t)|^2\,dt\right)^{1/2}}{\left(\int_c^dt^2\psi(t)|^2\,dt\right)^{3/2}}\right]
\label{eq:trueformulauntimate}
\end{eqnarray}

\noindent
(The constant $\frac{3+\sqrt{10}}{6\sqrt{2\pi}}$ is known to be the greatest lower bound \cite{Zolotarev1967}, and is conjectured to be also the least upper bound, but it is known \cite{KorolevShevtsova2010} only that the lub is at most 16\% greater than this constant.  This is sufficient.)

I now apply (\ref{eq:trueformulauntimate}) to the rate of pattern built-up in the double slit experiment.  The Schr\"odinger distribution for a double slit  is equation (16c) of (\cite{JenkinsWhite}, p. 313):

\begin{equation}
\left|\psi(t)\right|^2  = I_0\cos^2(n(t-\mu))\left(\frac{\sin(m(t - \mu))}{m(t-b)}\right)^2
\label{eq:Borndoubleslit}
\end{equation}	

\noindent
where $I_0$ is the height at distribution mean $t=\mu$, and

\begin{equation}
m  =\pi\left(\frac{w}{\sqrt{L^2 + (\mu-t)^2}}\right)\left(\frac{1}{\lambda}\right)
\label{eq:defintionofm}
\end{equation}

\noindent
with $w$ being the width of each slit, $L$ the distance from the two slits to the detector, and $\lambda$ the wavelength of the quantum particle, a photon or an electron.  In my analysis, the values of $w$, $L$, and $\lambda$ are taken from \cite{BachBatelaan2013}.

First divide the total one dimensional region $[a,b]$ into $i$ bins.  When there are $N$ total electrons detected, there will be $N_i$ electrons in the $i{\rm th}$ bin. The moments in (\ref{eq:trueformulauntimate}) must be central moments, which means that the origin of coordinates must be identical to the location of the mean.  For the double slit experiment, this mean and origin of coordinates is located at the center of the distribution (\ref{eq:Borndoubleslit}), which is to say, at the location of the central peak. In computing the RHS of (\ref{eq:trueformulauntimate}), I have rescaled (\ref{eq:Borndoubleslit}) and (\ref{eq:defintionofm}).

Inequality (\ref{eq:trueformulauntimate}) is for the time independent Schr\"odinger equation in for one electron in one spatial dimension.  The general case will be given elsewhere \cite{Tipler2019a}.

I compared formula (\ref{eq:trueformulauntimate}) with the data given in \cite{BachBatelaan2013} and illustrated in Figure \ref{fig:DoubleSlitDiffractionPatternbyElectrons} for nine values of $N$, namely $N = 13, \,54,\, 101,\, 200,\, 227,\, 302,\, 448,\, 613,\, {\rm and}\, 803$.  I always divided the entire interval into at least 10 bins, and occasionally less and more.   I allowed the bin labeling to start at $a$, and also at $b$. For all values for $N$, inequality (\ref{eq:trueformulauntimate}) holds, even if I let $\frac{3+\sqrt{10}}{6\sqrt{2\pi}}$ be 16\% greater.

\begin{figure}
\includegraphics[width=2.0in]{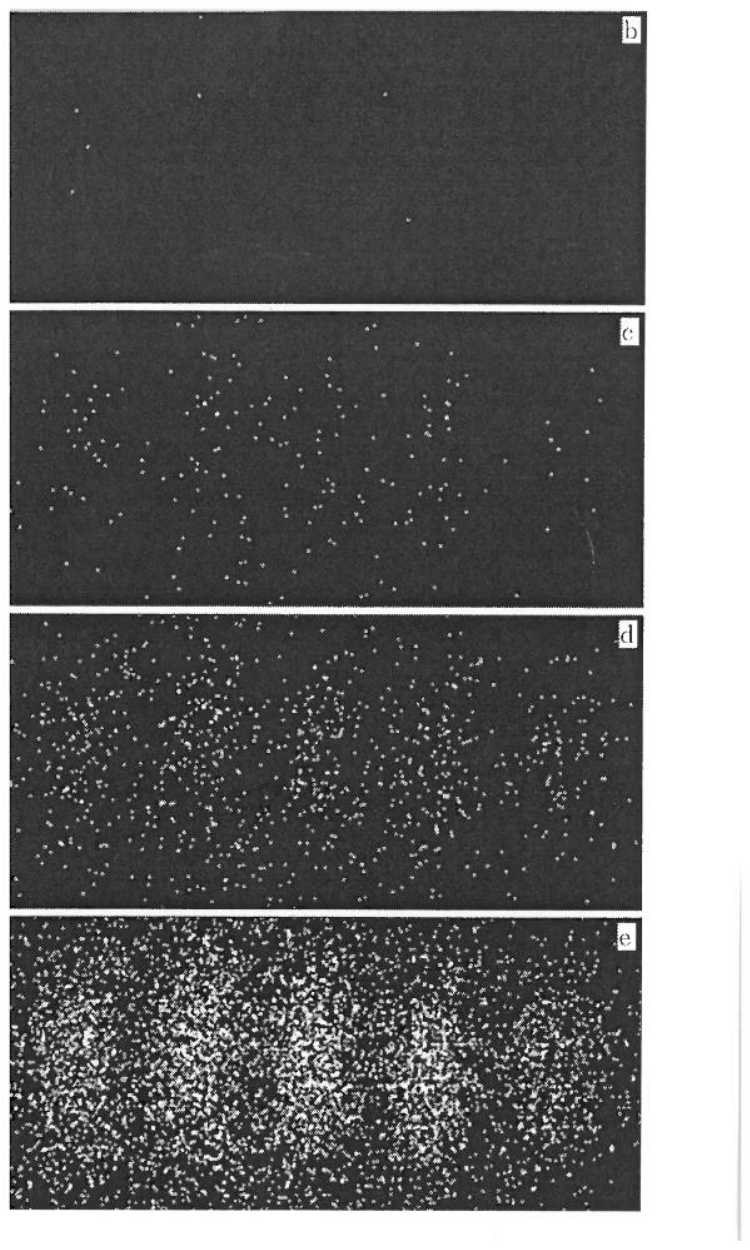}
\caption{Double Slit Diffraction Pattern by Electrons\label{fig:DoubleSlitDiffractionPatternbyElectrons}. The four images show the build-up of a double slit electron interference pattern. There are 7, 209, 1004, and 6235 electrons detected on the screen in (b)-(e) respectively.  (Figure taken from \cite{BachBatelaan2013}.)} 
\end{figure}

As the double slit experimenters \cite{BachBatelaan2013} do not give a statistical analysis of their experiment, neither can I.  However, statistical analysis is usually given to establish that one possibility is to be preferred over another.  In the double slit experiment, standard quantum mechanics {\it has} no theory of approach to the Born frequencies, as witness the fact that no such theory has ever been proposed, even though the double slit experiment and its pattern build-up has been in the physics textbooks for at least 70 years.  Many-Worlds quantum theory has such a theory, and it is manifestly confirmed by experiment.  As Rutherford once said, a good experiment establishes a fact without the necessity of statistical analysis \cite{Rutherford}.  Rutherford meant that an eyeball error estimate is sufficient.  This applies here.

In summary, I have shown in this paper that (1) the wave function $\psi$ is not a probability amplitude, but rather a relative density of universes in the multiverse amplitude; (2) that quantum mechanics is merely the most powerful form of classical mechanics, the Hamilton-Jacobi equation, taken seriously as being in one-to-one correspondence with physical reality --- all of its trajectories exist --- and required to be globally deterministic; (3) that the resulting Many-Worlds quantum mechanics is more general than standard quantum mechanics, since it includes the Born frequencies as a limiting case; (4) since Many-Worlds is more general, it has consequences not allowed by standard quantum mechanics; (5) one such consequence is a precise mathematical definition of a ``classical'' system; (6) that another consequence is it allows computation of the rate at which the Born frequencies are approached, and (7) the calculated rate agrees with experiment.

\end{document}